\documentclass[10pt, conference, compsocconf]{IEEEtran}
\ifCLASSINFOpdf
\else
\fi
\usepackage{graphicx}
\graphicspath{{./images/}}

\usepackage{amsmath,amssymb}
\begin{document}
%
\title{Graphical Analysis of Social Group Dynamics}


\author{\IEEEauthorblockN{Bogdan Gliwa, Anna Zygmunt, Aleksander Byrski}
\IEEEauthorblockA{Department of Computer Science\\
AGH University of Science and Technology\\
30-059 Krak\'ow, Poland\\
\{bgliwa,azygmunt,olekb\}@agh.edu.pl}
}
\maketitle

\begin{abstract}

Identifying communities in social networks becomes an increasingly important research problem. Several methods for identifying such groups have been developed, however, qualitative analysis (taking into account the scale of the problem) still poses serious
problems. This paper describes a tool for facilitating such an analysis, allowing to visualize the dynamics and supporting localization
of different events (such as creation or merging of groups). 
In the final part of the paper, the experimental results performed using the benchmark data (Enron emails) provide an insight
into usefulness of the proposed tool.

\end{abstract}


%
\IEEEpeerreviewmaketitle

\section{Introduction}
Current trends in identification of groups in complex network analysis tend to go beyond static analysis (see, e.g., \cite{palla2007,Spiliopoulou2011}) and 
take into account the dynamic character of the environment, mostly concerning the quantitative analysis of such dynamic groups. Qualitative analysis becomes a very difficult task, due to huge network sizes, possible number of groups and time-dependence. In this paper, GEVi (Group Evolution Visualisation)---a tool for the graphical analysis of the evolution of groups will be presented.

Real-life networks are characterized by rapid changes and the groups that may be located are mostly short-lived and elusive. In order to analyse certain processes or trends occurring in groups, different time periods should be taken into account.
Observation of changes should lead into stating the reasons for creation, extension or disappearance of certain groups.
It is to note, that an additional challenge is the fact, that one user may be a member of many groups.
Correlating of the observation of the network dynamics with external events may lead to explaining of certain processes
occurring in the structure of groups and to allow prediction of future events.

In the paper, after presenting the state of the art and describing the utilized method of groups extraction, the features
of the presented tool are shown and the experimental results obtained from popular Enron dataset are discussed.

\section{Related Work}
Initially finding groups (communities) in large social networks was made possible by extracting certain features from the network and analyze them
on higher level of abstraction: the network could be represented in an equivalent, but much less complex form as groups and the relationships between them \cite{wasserman1994}. Nowadays, group finding techniques allow not only to simplify the network, but moreover, to analyze certain  processes in micro and macro scale.
There are many definitions of a group, but usually it is assumed that the group is a set of vertices which communicate to each other more frequently than with vertices outside the group. Many methods of finding groups (mainly in static graphs) have been proposed \cite{Fortunato2010}. Nowadays, many results regarding the the dynamics of the network, taking into account the time and
its impact on the life cycle of the groups are published \cite{Asur2009} \cite{Spiliopoulou2011}.
Palla et al. in \cite{palla2007} identified basic events that may occur in the life cycle of the group: growth, merging, birth, construction, splitting and death. They did not give any additional conditions. Asur in \cite{Asur2009} introduced formal definitions of five critical events. Greene in \cite{Greene2010} presented
a review of the fundamental events describing group evolution and formulated these key events in terms of rules.

In \cite{reda2011}, a tool for visualization of the evolution for non-overlapping groups was proposed. With this tool one can analyse the membership of certain individuals in the group, rather than the evolution of the group itself.

\section{The method of groups extraction in dynamic environment}
We have used SGCI (Stable Group Changes Identification) algorithm and CPM (Clique Percolation Method) \cite{palla2005} as a group extraction method. The algorithm consists of four main steps: identification of  short-lived groups in each separated time interval; identification of  group continuation (using modified Jaccard measure), separation of the stable groups (lasting for a certain time interval) and the identification of types of group changes (transition between the states of the stable group). A detailed description of the algorithm is in \cite{gliwa2012}.

We used the set of events identified in \cite{gliwa2012}, applying more general methods for their identification. The algorithm identifies transitions between groups observed at time $t$ and the groups observed at the time $t+1$ (their successors). This is achieved by comparing the size of the source groups, with each of their successors, rather than the difference in size between all successors. 

\begin{figure*}[ht]
\centering
\includegraphics[scale=0.7]{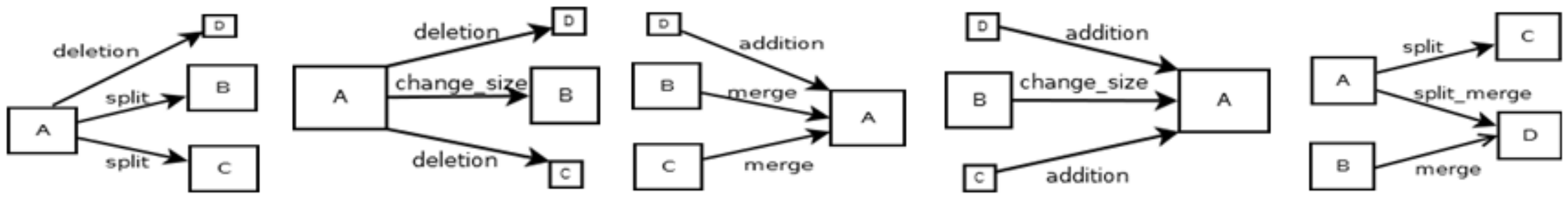}
\caption{Illustration of events.}
\label{fig:events}
\end{figure*}

For various reasons, it is interesting to observe lifespan of communities. How social network is evolving? What are the reasons for appearance of communities in social network, how they grow or shrink, what are the causes of new members joining and abandoning the old? Whether the community observed in two time periods is the same community, even though, for example, there is no common members?
 
There are many interesting questions, but the available tools lack possibilities of simple, preferably graphical, analysis of groups life-cycle.
A tool that may be used both for quantitative and qualitative analysis presenting graphical visualization of events and changes in the network would be much desired.

\section{Tool for graphical analysis of network evolution (GEVi)}
The GEVi visualizes groups in timeslots and displays transitions between them in a form of graph. Each distinct hierarchy of group evolution is displayed as a separate graph.
To implement visualisation we used JGraph\footnote{http://www.jgraph.com/jgraph.html} Java-based library. 

\subsection{Visualisation technique}
The groups and transitions between them are represented using hierarchical (Sugiyama type) layout. It \cite{bastert2001} has several interesting features: there are few edge crossings, the nodes are evenly distributed and the edges are as straight as possible.
The Sugiyama layout is a method for visualizing directed graphs and consists of the following stages:
\begin{itemize}
   \item cycle removal -- some edges are reversed in order to make the graph acyclic (at the end of algorithm they are reversed again to initial state),
	\item layer assignment -- assignment of the vertices to layers (if there are edges that pass not only through adjacent layers, the dummy vertices are introduced),
	\item crossing reduction -- in each layer the ordering of vertices is calculated in order to minimize the number of edge crossing,
	\item coordinate assignment -- positioning of vertices so they do not overlap each other and that vertices not lie on the straight lines between two adjacent vertices from different layers, placing edges.
\end{itemize}

In our case, the transitions between groups cannot form cycles in graph so we omitted first stage. The second stage was simple in our situation because the groups are assigned to timeslots where they were extracted. As  the layers in the graph represent the timeslots, so we preassigned nodes in the graph to their layers.
For reduction of crossings and coordinate assignment, some variants of median method described by Gansner \cite{Gansner1993} were used. 

\subsection{Features}
\begin{figure}[ht]
\centering
\includegraphics[scale=0.7]{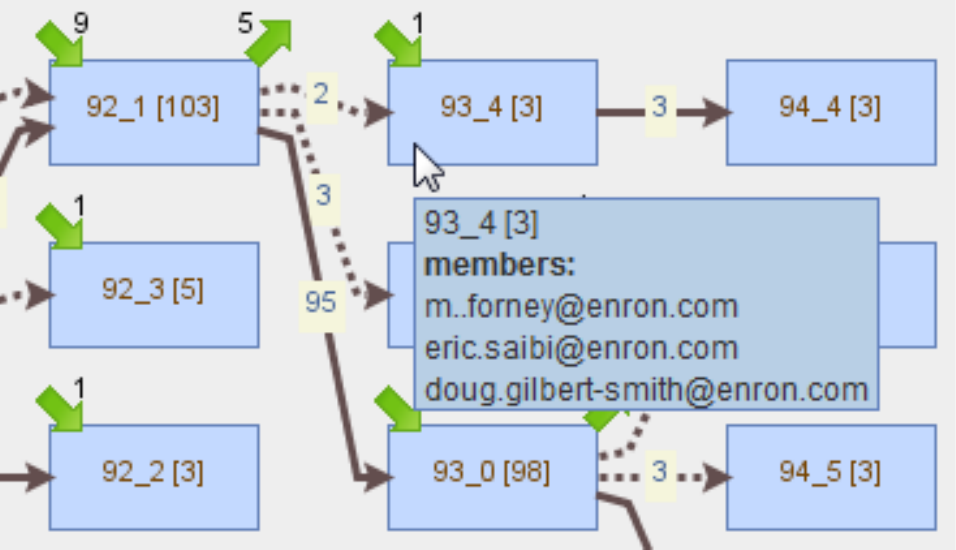}
\caption{Visualisation with showing context menu for group.}
\label{fig:screen1}
\end{figure}

In GEVi, each group is labelled in a form $timeslotNumber\_groupNumber$ which eases the identification of the groups during their evolution.
GEVi enables not only analysis of transitions between groups in different time slots (fig. \ref{fig:screen1}) but also shows the size of groups (in square brackets inside vertices), denoting how many members get inside the group during each group transition (label on transition) and how many of them get outside during each group transitions (in a form of number close to the green arrow---the green arrow pointing in the direction of the top-right corner stands for the number of members that go outside groups connected by outgoing transitions and the green arrow pointing in the direction of the bottom-right corner stands for the number of members that go into given group). For instance, the group 92\_1 from fig. \ref{fig:screen1} has 2 input edges (96 members flow from predecessors of that group to the given one) and additionally 9 members (not belonging to predecessors of that group) come to this group. The group has 3 outgoing edges (100 members flow to its successors) and additionally 5 members leave that group.

Some transitions are displayed as dashed arrows---this indicates that groups between given transition differ significantly in size (one of them is at least 10 times bigger than the second one). Such transitions represent events described as addition or deletion (depending whether small group attaches to the larger or small group detaches from the larger one).

In the transition pop-up menu, there is an additional information about stability during group transition and in the group pop-up menu (fig. \ref{fig:screen1}) - the members of the group are listed.

\begin{figure}[ht]
\centering
\includegraphics[scale=0.7]{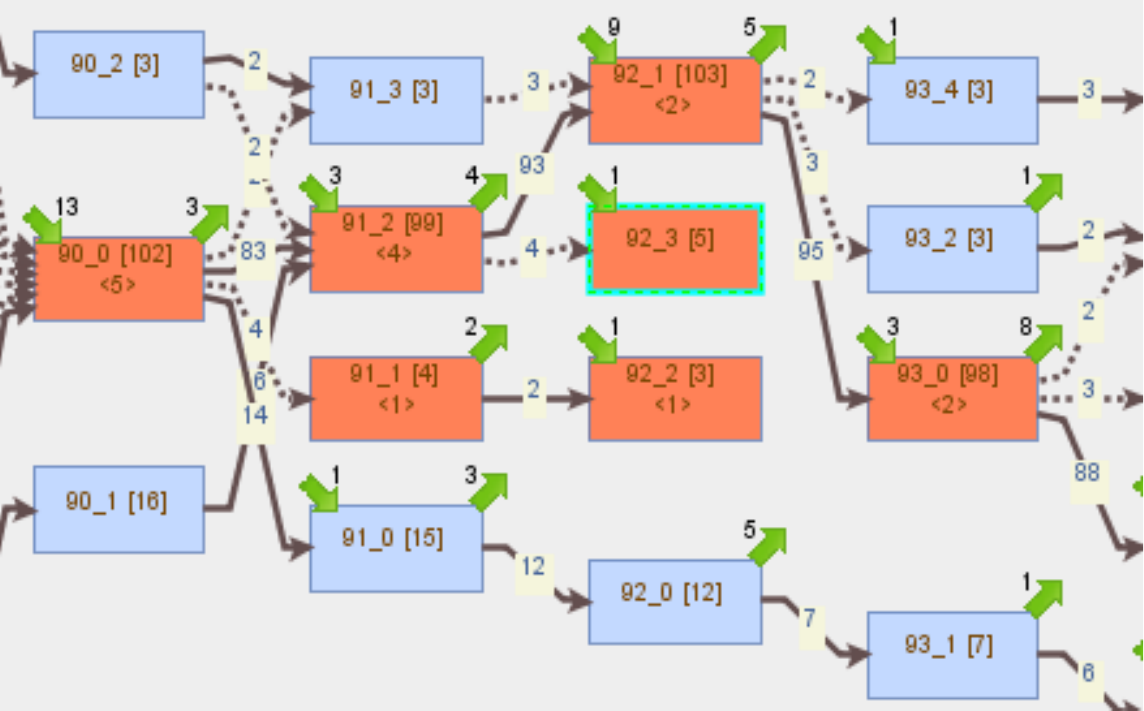}
\caption{Visualisation showing common members for group 92\_3.}
\label{fig:screen2}
\end{figure}

GEVi also gives us information about overlapping of the members between the groups.
After selecting of the group, all other groups that have in common at least one member with the selected one are highlighted (fig. \ref{fig:screen2}) and the information is displayed, regarding the number of common members (number between characters $<$ and $>$ inside vertex) and in the pop-up menu the members of all highlighted groups common with the selected one are shown. 

To be more useful, GEVi supports also zooming graphs and searching for groups by its name in a form of $timeslotNumber\_groupNumber$ (after finding the group, the focus is set and the view is centered).

\subsection{Model}
In this section a simple model for describing the analysis of the network dynamics is proposed.

A complex network or social network may be of course described using standard definition of 
a graph:
\begin{equation}
N = \langle V, E \rangle
\end{equation}
where:
$V \subset \mathbb{N}$, stands for a finite set of vertices, 
that is:
\begin{equation}
V = \{i: i\in \mathbb{N} \land i<i_{max}\}
\end{equation}
and $E=\subset V \times V$ is a finite set of edges. 

Striving to provide means for observation of groups that are formed in a certain time moment,
let us consider the following space of system states: $G = 2^V$.
The elements of $G$ are any possible subsets of $V$. Now, observing the system in a certain time
moment, it may be seen that the set of vertices is decomposed into following subsets:
\begin{equation}
G \ni g_t = \{ g_{t,k} \}, t,k \in \mathbb{N}.
\end{equation}
each subset may be described as:
\begin{equation}
g_{t,k} = \{v_1, \ldots, v_{max_{t,k}}\}.
\end{equation}
where $max_{t,k}$ stands for maximum number of the individuals in the group.
Note, that the subsets observed at certain time $t$ may contain the
same elements (they may overlap).

Now, let us define the graph depicting the dynamics of the complex network. Again, as it is a graph,
the definition is similar to the classical one:
\begin{equation}
D = \langle V_D, E_D \rangle
\end{equation}
where: $V_D=(t,k) \in \mathbb{N} \times \mathbb{N}$, and
$E_D=V_D \times V_D$
so this graph is composed of labels utilized before, in the definition of the complex network and the
groups. Note, that this definition spans to the whole observation time of the network.

The above-presented simple formalism is aimed to ease the definition of observed events and other primitives.

For example, let us define Modified Jaccard measure
\begin{equation}
MJ(A,B)=\begin{cases}
		0, & \text{if $A=\emptyset \vee B=\emptyset$,}\\
	  max(\frac{|A \cap B|}{|A|},\frac{|A \cap B|}{|B|}), & \text{otherwise.}
	\end{cases}
\end{equation}
and ratio of groups size
\begin{equation}
ds(A,B)=max(\frac{|A|}{|B|},\frac{|B|}{|A|})
\end{equation}
where $A \neq \emptyset \wedge B \neq \emptyset$.

Transition $t_{g_{i,k},g_{i+1,l}}$ can be defined as:
\begin{equation}
t_{g_{i,k},g_{i+1,l}}: \exists g_{i,k} \wedge \exists g_{i+1,l} \wedge MJ(g_{i,k},g_{i+1,l})\geq th
\end{equation}
where $th$ means threshold for creation of transition (in experiments we set value of $th$ to 0.5).

Due to the limited space in this article we present only formulation for split\_merge event (figure \ref{fig:events} shows illustration for most events):

The event \emph{split\_merge} occurs when group $g_{i,k}$ divides into 2 or more groups in next time slot, these groups from next time slot have similar size to $g_{i,k}$, the group $g_{i+1,l}$ is created from 2 or more groups from previous time slot and these groups from previous time slot have similar size to $g_{i+1,l}$
\begin{eqnarray}
& t_{g_{i,k},g_{i+1,l}}: ds(g_{i,k},g_{i+1,l}) < sh \wedge \nonumber \\
& [\exists t_{g_{i,m},g_{i+1,l}}: m \neq k \wedge ds(g_{i,m},g_{i+1,l})< sh] \wedge \nonumber \\
& [\exists t_{g_{i,k},g_{i+1,n}}: n \neq l \wedge ds(g_{i,k},g_{i+1,n})< sh] 
\end{eqnarray}

where $sh$ stands for the  threshold for ratio of groups size, which in experiments was set to 10. 

\section{Graphical analysis of Enron dataset}

\subsection{Dataset}
We analyzed one of the most popular datasets in complex network analysis: Enron emails. The dataset was prepared in the form of MySQL database and described by Shetty and Adibi\cite{shetty2004eed}. They made it publicly available \footnote{http://www.isi.edu/$\, \tilde{\; }$adibi/Enron/Enron.htm}.

The analyzed data contains emails from 151 users and 252~759 messages from the following period of time: 5.01.1998-3.02.2004. Some messages were sent to group of people, therefore  
such messages can be expanded into multiple messages between single sender and single recipient. The database contains 2~064~442 of such expanded messages. 
We restricted messages to the ones that were exchanged only between employees (that were listed in the database in separate table). After rejection, there were 50~572 left of the expanded messages.

\subsection{Group extraction and evolution}
The analyzed period was divided into time slots, each lasting 30 days. The neighbouring slots overlap each other by 50\% of their duration and in the examined period of time there are 149 time slots.

After separation of time slots we extracted the groups in each time slot. We used CPM method of community extraction (CPMd version from {CFinder\footnote{http://www.cfinder.org/}} tool)
for k=3. For this parameter groups were extracted in slots between 31 (15.04.1999-15.05.1999) and 108 (13.06.2002-13.07.2002)---in other time slots there were so few messages between users that no groups were formed (for higher k values the range of time slots containing any groups is even more narrow).

Transitions between groups were assigned using our method SGCI described earlier. The threshold on modified Jaccard measure was set on level equals 0.5.

\subsection{Group sizes}
Running simple statistic algorithms we determined that most of the groups are small---groups that have their size equal 10 or less constitute about 80\% of all groups.

Using GEVi, we can observe the size for each group as it was demonstrated on fig. \ref{fig:numbersComparison}. For instance, the group 92\_1 has 103 members and size of group 93\_4 equals 3.

\subsection{Number of groups in timeslots}
\begin{figure}[ht]
\centering
\includegraphics[scale=0.55]{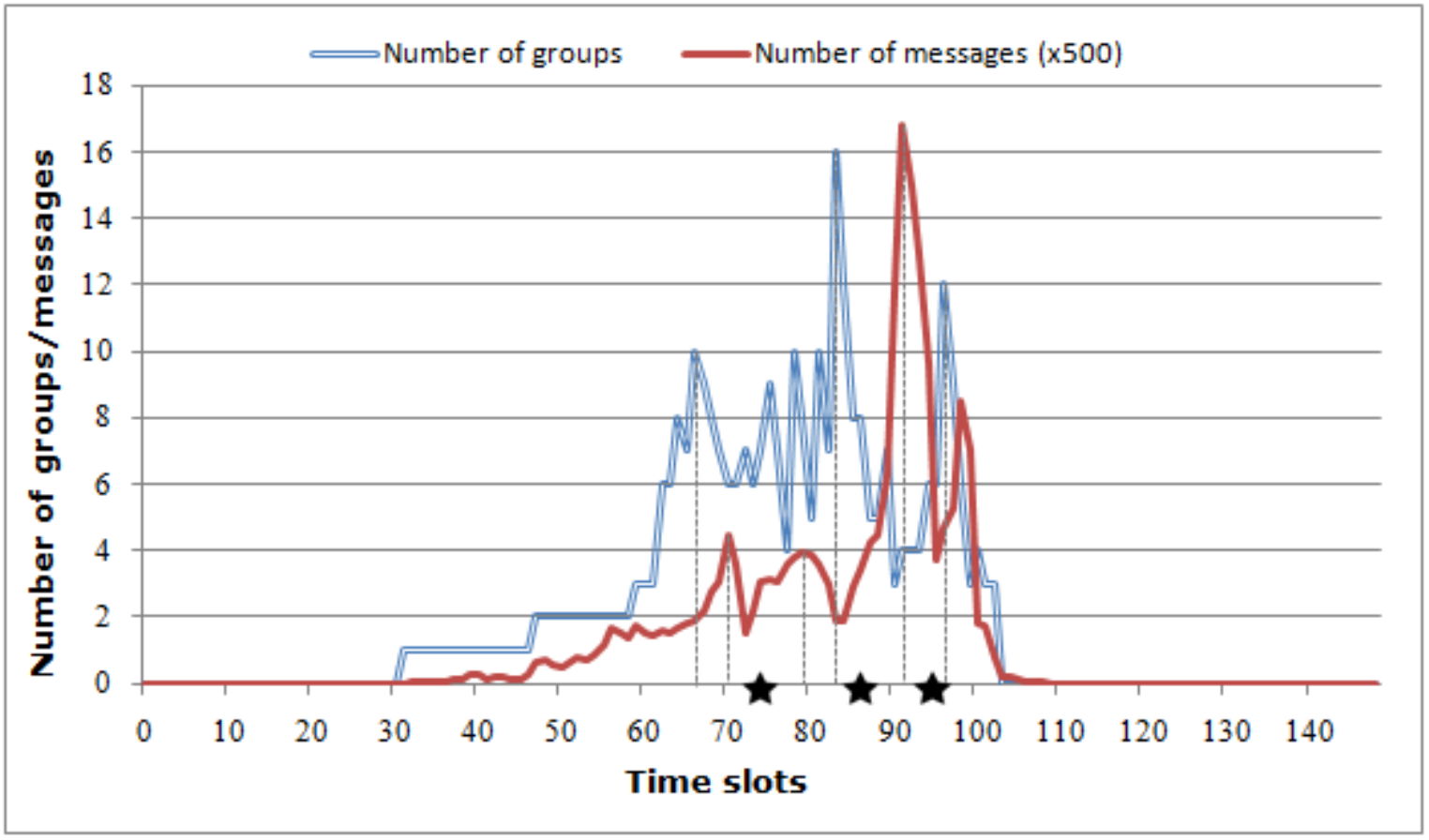}
\caption{Number of groups and messages in time slots.}
\label{fig:numbersComparison}
\end{figure}

Fig. \ref{fig:numbersComparison} shows the number of groups and messages in time slots. The stars on chart represent key events from timeline of Enron:
\begin{itemize}
    \item 12.02.2001 - Skilling is named CEO (slots 74, 75),
	\item 14.08.2001 - Skilling resigns as CEO (slots 86, 87),
	\item 2.12.2001 - Enron files for bankruptcy (slots 94, 95).
\end{itemize}

We were inspired by work of Collingsworth, Menezes and Martins \cite{Menezes2009}, who also analyzed Enron dataset and in the cited paper, there is presented a chart showing the relation between the number of emails sent by users and key events for company (the same as we recalled above). They noticed that peaks of the exchanged emails happened before key real events at an average of 2 months earlier. Therefore, we prepared similar chart as they used in their work---the chart presenting number of messages (we are showing only number of messages exchanged between employees) in time.

We can compare these 2 charts in fig. \ref{fig:numbersComparison} and we can observe that peaks on chart with number of messages precede mentioned events but on chart with number of groups in 2 first cases peaks precede events and the last peak is right after the last event.

The number of groups in each time slot can be easily perceived---the groups from the same time slot in the same hierarchy are positioned vertically one above the other.

\subsection{Stability of groups in timeslots}

\begin{figure}[ht]
\centering
\includegraphics[scale=0.55]{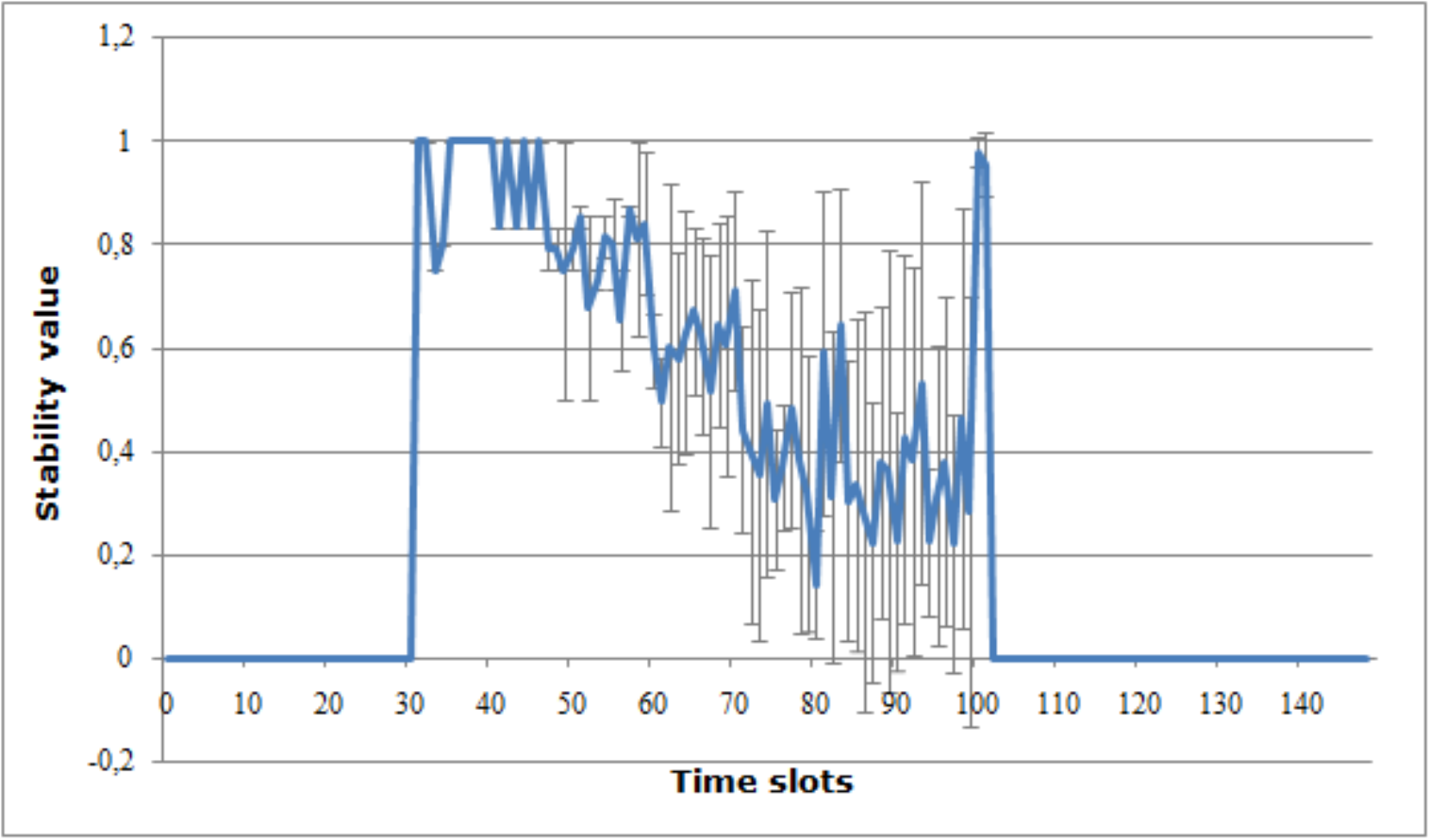}
\caption{Stability of groups in time slots (mean and standard deviation).}
\label{fig:stabilitySlots}
\end{figure}

In fig. \ref{fig:stabilitySlots},  mean stability (with standard deviation) between groups in slots is presented (e.g., stability in the slot 100 corresponds to stabilities between groups from the slot 100 and the slot 101). We can observe that stability gradually decreases in time until slot 100 (13.02.2002-15.03.2002) which happened about 3 months after bankruptcy of Enron. We can also notice that when the mean stability decreases, the standard deviation has large values, which is caused by many deletions and additions.

\begin{figure}[ht]
\centering
\includegraphics[scale=0.75]{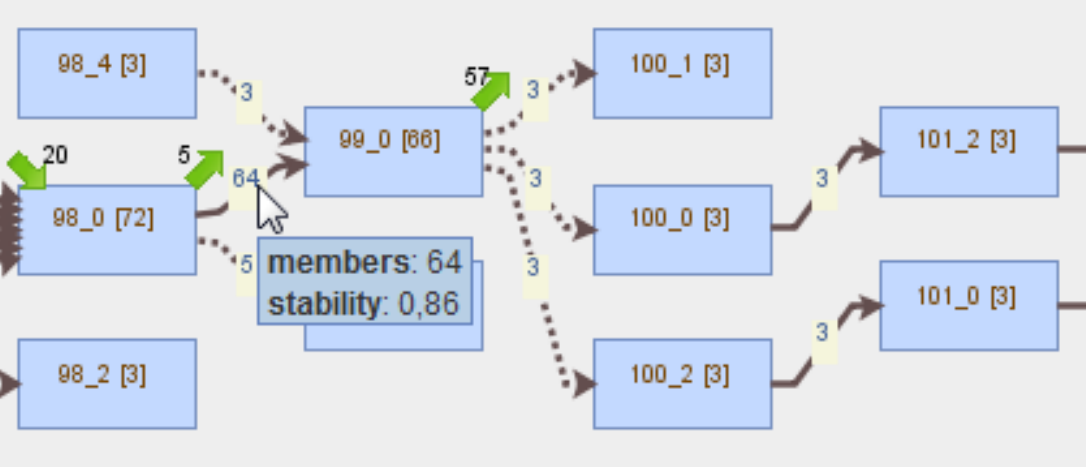}
\caption{Stability for chosen transition on visualisation.}
\label{fig:stabilityVis}
\end{figure}

The stability of each transition between groups can be observed in GEVi when hovering mouse pointer over a certain chosen group---see fig. \ref{fig:stabilityVis},
or indirectly: if in a given time slot there are more dashed transition arrows, the mean stability is expected to be less than in timeslots when there are mainly solid arrows, which is presented in fig. \ref{fig:stabilityVis} (mean stability between groups in slots 99 and 100 is less than between groups from slots 100 and 101).

\subsection{Exchange of members of group in time}
\begin{figure*}[ht]
\centering
\includegraphics[scale=0.6]{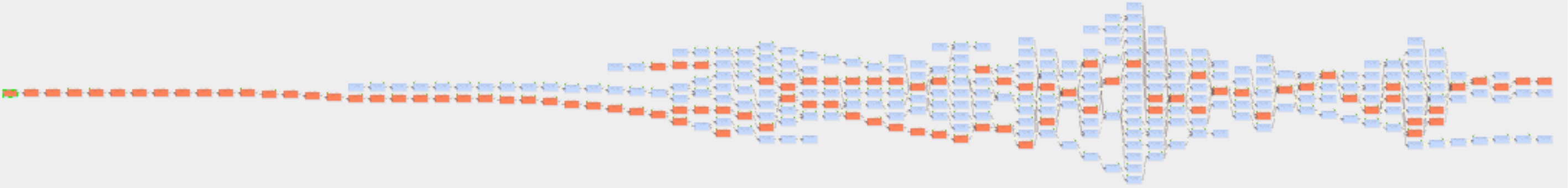}
\caption{Visualisation of groups that have common members with first group in hierarchy.}
\label{fig:coreVis}
\end{figure*}

Four different hierarchies can be visualised in GEVi. The most interesting one is shown in fig. \ref{fig:coreVis}, where the highlighted groups are the ones having in common at least one member with the first group in this hierarchy (group labelled as 31\_0). The mentioned group has 3 members and as we can notice, in each next time slot (every time slot has different vertical layer in visualization) there is at least one group that has any common members with that group (what is presented in fig. \ref{fig:coreVis}). In the last time slot for this hierarchy (slot 102) the only one person from the initial group is present. 

This example shows how this tool can be used in analyzing, how long a given group can exist without complete exchange of initial members of group.

\subsection{Common members between groups in the same time slot}

The maximum number of common members between each group pair from the same time slot equals 3. It seems that in about 22\% of all pairs from the same time slot there is at least one common member.

GEVi makes possible checking common elements for each selected group with the other ones. For instance, in fig. \ref{fig:screen2} we can see that group 92\_3 has 5 members and with group 92\_1 has 2 common members, with 92\_2 has 1 common member and there is no common members with group 92\_0.

\subsection{Overlapping groups in the same time slot}

Most groups overlap at least with one another group in the same timeslot. The groups that do not overlap with others constitute about 30\% of all groups.

The presented tool enables possibility to check the group overlapping in the same time slot. Referring to fig. \ref{fig:screen2}, one can see that the group 92\_3 overlaps with 2 other groups in the same time slot.

\subsection{Analysis of behavior of group dynamics close to Enron bankruptcy}
\begin{figure*}[ht]
\centering
\includegraphics[scale=0.72]{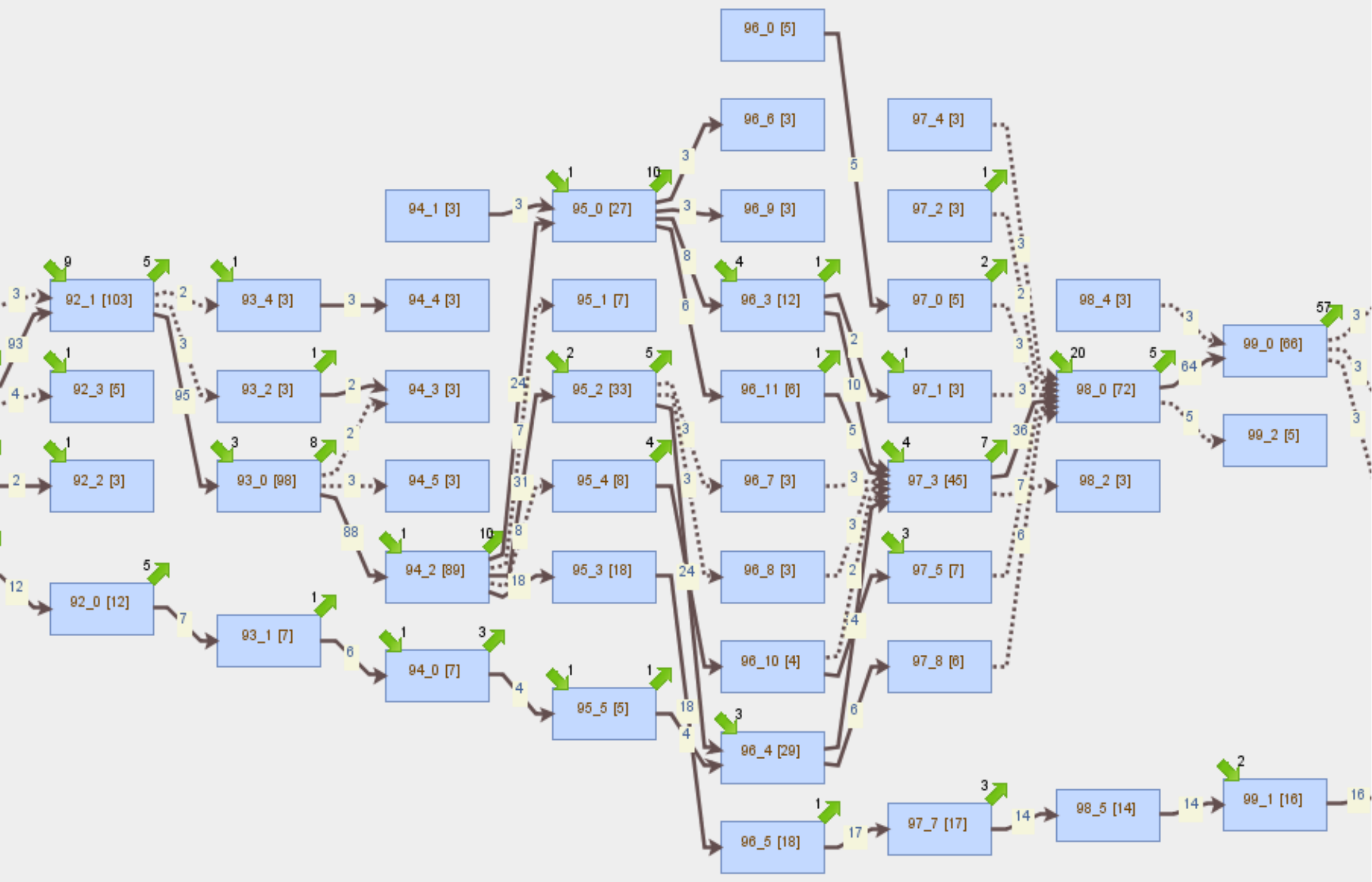}
\caption{Group dynamics close to Enron bankruptcy.}
\label{fig:example}
\end{figure*}

Enron bankruptcy took place in 94th and 95th time slots (slots are overlapping). Fig. \ref{fig:example} shows that right before that event, there exist several small and one big group in time slots, but in 95th time slot large group 94\_2 splits into some smaller groups. It could suggest that people were afraid of their situation and prefer leading communication with subgroup of people who seemed to be more trustworthy. We can also observe that after some time the situation changes and again most people belongs to one large group. 
Another interesting remark is that in 96th time slot (just after Enron bankruptcy), there is a peak of group numbers that  could also imply certain doubtfulness of people about their situation and interacting with other people (via mail) in small groups. 

\section{Conclusion}
In this paper GEVi's features were described. The tool allows to construct higher abstraction level charts 
and use them for visualization of certain group events. In the future we plan to add possibilities of detecting
new events and to employ different benchmark and real-world data to tune-up the proposed network analysis tool.

{\bf Acknowledgments.}
The work was supported by Research project No. O ROB 0008 01
,,Advanced IT techniques supporting data processing in criminal analysis'', funded by the Polish National Centre for Research and Development.

\bibliographystyle{IEEEtran}
\bibliography{cason,groups}

%
%
%

\end{document}